\documentclass[compsoc, conference, letterpaper, 10pt, times]{IEEEtran}
\usepackage{grumble}
\usepackage{subfig}
\usepackage{color}
\usepackage{epsfig,endnotes}
\usepackage{graphicx}
\usepackage{url}
\usepackage{multirow}
\usepackage{booktabs}
\usepackage{amsmath}
\usepackage{balance}
\graphicspath{ {figs/} }
\usepackage{enumitem}
\usepackage{pifont}
\setlist{nosep}
\usepackage{tikz}
\usepackage{float}
\usepackage{balance}
\usepackage[nocompress]{cite}

\newcommand{\subsubsecspace}{\vspace{0.05in}}

\newcommand{\subsubsectitle}[1]{\subsubsecspace\noindent\textbf{#1}}
\newcommand{\eg}{{\em e.g.,\ }}
\newcommand{\ie}{{\em i.e.\ }}

\ifCLASSOPTIONcompsoc
  \usepackage[nocompress]{cite}
\else
  \usepackage{cite}
\fi
\ifCLASSINFOpdf
\else
\fi
\begin{document}
%
\title{I Spy with My Little Eye: Analysis and Detection of Spying Browser Extensions}



%
\author{\IEEEauthorblockN{Anupama Aggarwal\IEEEauthorrefmark{1},
Bimal Viswanath\IEEEauthorrefmark{2},
Liang Zhang\IEEEauthorrefmark{3}, 
Saravana Kumar\IEEEauthorrefmark{4}, \\
Ayush Shah\IEEEauthorrefmark{1} and
Ponnurangam Kumaraguru\IEEEauthorrefmark{1}}
\IEEEauthorblockA{\IEEEauthorrefmark{1}IIIT - Delhi, India, email: anupamaa@iiitd.ac.in, ayush13027@iiitd.ac.in, pk@iiitd.ac.in}
\IEEEauthorblockA{\IEEEauthorrefmark{2}UC Santa Barbara, email: viswanath@cs.ucsb.edu}
\IEEEauthorblockA{\IEEEauthorrefmark{3}Northeastern University, email: liang@ccs.neu.edu}
\IEEEauthorblockA{\IEEEauthorrefmark{4}CEG, Guindy, India, email: savkumar90@gmail.com}}


\maketitle

\begin{abstract}
In this work, we take a step towards understanding and defending against \textit{spying browser extensions.} These are extensions repurposed to capture online activities of a user and communicate the collected sensitive information to a third-party domain. We conduct an empirical study of such extensions on the Chrome Web Store. First, we present an in-depth analysis of the spying behavior of these extensions. We observe that these extensions steal a variety of sensitive user information, such as the complete browsing history (e.g., the sequence of web traversals), online social network (OSN) access tokens, IP address, and geolocation. Second, we investigate the potential for automatically detecting spying extensions by applying machine learning schemes. We show that using a Recurrent Neural Network (RNN), the sequence of browser API calls made by an extension can be a robust feature, outperforming hand-crafted features (used in prior work on malicious extensions) to detect spying extensions. Our RNN based detection scheme achieves a high precision (90.02\%) and recall (93.31\%) in detecting spying extensions.

\end{abstract}



%
\IEEEpeerreviewmaketitle

\section{Introduction}

\noindent
Many websites today rely on advertising to sustain their services.
Online ad platforms have developed sophisticated user tracking systems to capture user behavior across websites in order to generate highly personalized ads. Several studies have been conducted to understand \textit{third-party user tracking on the web}, where a third-party entity (like social media widget), embedded by the first-party site the user visits, can log the presence and other activities of the user on that site~\cite{englehardt2016online}\cite{roesner2012detecting}. Unfortunately, user tracking has reached to a state where even privileged software running on a user's browser is stealing user information~\cite{starovextended}. In this work, we take a step towards understanding and defending against this new type of privacy risk---{\em browser extensions that spy on user information}.

A browser extension is an add-on software that enhances functionality of the browser. We consider a spying extension as one which accesses user information (\eg browsing history) and sends the information to third-party domains, when its core functionality does not require such information communication (Section~\ref{sec:dataset}). Such information theft is a serious violation of user privacy. For example, a user's browsing history can contain URLs referencing private documents (\eg Google docs) or the URL parameters can reveal privacy sensitive activities performed by the user on a site (\eg online e-commerce transactions, password based authentication). Moreover, if private data is further sold or exchanged with cyber-criminals~\cite{Lifehacker:pi}, it can be misused in numerous ways. Left unchecked, users may be exposed to various risks, including identity theft~\cite{krishnamurthy2006generating}, and financial loss.

Compared to other types of user tracking, spying by browser extensions deserves special attention. Unlike web pages, browser extensions persist through the entire browser life cycle, and can access in-depth user browsing activities across all websites. By making privileged API calls to the browser, extensions can access the user's browsing history, current open tabs, record keyboard inputs and more. On the other hand, third-party web trackers can only track users on sites where they are embedded by the publisher, and thus obtain a ``limited'' or fragmented view of a user's entire browsing behavior. Also note that extensions are available on most popular browsers (\eg Chrome, Safari, and Firefox) and have 510 million cumulative installs in Chrome. Therefore, spying extension developers have a large user base that they can target.

Identifying spying behavior is challenging. Given an extension, automatically identifying the flow of sensitive user information is a non-trivial task~\cite{bandhakavi2011vetting}. Moreover, we observe spying extensions that obfuscate user information before sending it to remote servers, and those that switch between spying and non-spying states during their life-time (\eg using remote Command-and-Control triggers). Such behavior further complicates the identification.

To bootstrap our analysis, we resort to expert manual investigation of behavioral logs (traces of network, storage, and browser API requests made by an extension) of extensions to identify spying behavior (Section~\ref{sec:dataset}). 
To reduce manual effort, we apply several heuristics to automatically short-list a candidate set of possible spying extensions, among all the 43,521 extensions on the Chrome store. After manually investigating over 1,000 extensions in the candidate set, we discover 218 spying extensions. This dataset provides an opportunity to better understand spying behavior and to build a defense scheme that can automatically detect spying extensions (ie without any manual effort). 

We start by uncovering the modus operandi of spying extensions and analyze their characteristics (Section~\ref{sec:analysis}). Spying extensions steal browsing history, social media access tokens, IP address and geolocation of users. Surprisingly, spying extensions are as popular (based on user base) and have similar ratings as other extensions on the Chrome store. We suspect that users are mostly unaware of the spying behavior as only 12 out of 218 extensions received reports of any suspicious behavior from users. 

Next, we focus on automatically detecting spying extensions. While a lot of prior work has focused on HoneyPages to trigger and catch malicious behavior by extensions~\cite{kapravelos2014hulk}, and information flow control based approaches~\cite{bandhakavi2011vetting}\cite{dhawan2009analyzing}\cite{jang2010empirical}; we take an alternate approach of leveraging recent advances in machine learning (ML) for detection. Practicality and effectiveness of ML-based defenses has encouraged the industry (\eg Google) to also adopt such approaches to protect against malicious extensions~\cite{jagpal2015trends}. 

As is the case with any ML-based defense, we start by first identifying a robust feature. Prior work on detecting malicious extensions relies on extensive feature engineering to craft a large number of features (\eg features  based on extension meta-data, source code, and network, storage and API requests)~\cite{jagpal2015trends}. Surprisingly, our analysis reveals that most of the existing features are ineffective in detecting spying extensions (Section~\ref{subsec:prior}). Instead, browser API calls modeled as sequential data provides the highest predictive power, thus obviating the need for complex hand-crafted features. We further discuss the benefits and robustness aspects of using API call sequence for detection. 

However, browser API call sequences can contain complex sequential patterns which makes traditional n-gram based sequence classification approaches ineffective in our case. We observe that recent developments in a class of deep neural networks called Recurrent Neural Networks (RNN) are well suited for our scenario. We present an RNN based detection scheme that is capable of learning sophisticated patterns in API call sequences~\cite{de2015survey}. An in-depth evaluation of our RNN based approach shows that it outperforms traditional ML schemes and achieves a high precision (90.02\%) and recall (93.31\%) in detecting spying extensions (Section~\ref{subsec:rnn}). Further, our RNN-based classifier additionally identified 65 previously unknown spying extensions. Finally, we discuss deployment aspects of our approach. While our scheme can be used in a centralized setting by analyzing the extensions in a controlled environment, we also explore the potential for pushing the detection to the edge at user's browser (Section~\ref{sec:edge}).

\section{Background} 
\label{sec:background}
\noindent We provide a brief overview of the building blocks of a Google Chrome extension and how developers can publish extensions for end-users~\cite{Google:bs}. Although we focus on Chrome extensions, most of the concepts apply to other browsers as well, for example, Firefox with WebExtensions~\cite{Mozilla:2017yt}. 

\subsubsectitle{\textbf{Extension Architecture. }} Browser extensions are software programs running on web browsers, providing extended browsing functionality to users. An extension consists of JavaScript, HTML, CSS, and other web resources that are needed for rendering web pages. 
The main logic of an extension is usually placed in a \emph{background page} where programs keep running during the entire browser session. Many extensions also use \emph{content scripts} that are programs injected into a web page to gain access to the page resources. 

Similar to web pages, extensions have access to all Web APIs, such as standard JavaScript API, XMLHttpRequest, and DOM. More importantly, Chrome extensions are exposed to a set of privileged browser APIs, called the \emph{Chrome API}. Based on the functionality, Chrome API is bundled as various feature endpoints, such as web requests (e.g., listening or tampering network requests), bookmarks, history, cookies, storage, and tabs.
In order to gain access to the Chrome API, extension developers have to specifically request permissions in the \emph{Manifest file}. When installing an extension, users have to decide to accept these permission requests or not. 
For example, the \emph{cookie} permission in the Manifest file enables extensions to make API calls to read cookies (\emph{cookie.get}) and change cookies (\emph{cookie.set}). With the necessary permissions, extensions can monitor all user activities in browser. 
It should be noted that other browsers, Safari and Firefox also provide a similar privileged Browser API to extensions~\cite{Mozilla:2017yt}\cite{Apple:2017ht}.

\subsubsectitle{\textbf{Chrome Web Store. }} The Chrome Web Store serves as a repository for all extensions and provides a low entry barrier for developers to publish extensions. However, due to the rampant increase of malicious extensions, Chrome now asks for a one-time sign-up fee to publish extensions~\cite{Chrome:fk}, and does not allow users to install extensions sourced outside the Chrome Web Store ecosystem~\cite{Chrome:2015rz}, unless a deliberate developer mode is enabled by the user. Additionally, Google is known to constantly monitor the extension store to identify malicious browser extensions~\cite{jagpal2015trends}.
Despite these countermeasures, our findings show persistent presence of spying extensions on the Chrome Web Store.

\section{Data Collection} \label{sec:dataset}

\subsection{Ground Truth Data of Spying Extensions}
\noindent To conduct our study, we first need a large sample of spying extensions. In this section, we describe our methodology for obtaining a ground-truth dataset of spying extensions. In the rest of the paper, we use this dataset to uncover the operational aspects of spying extensions and to propose a machine learning-based technique to automatically detect spying extensions.

\subsubsectitle{\textbf{Spying Extensions:} } We define a \textit{spying} extension as one which accesses and sends user information to a third-party domain, when the claimed functionality of the extension (on the Chrome store) does not mention the requirement of such user information. Leakage mechanisms include communication events based on Web API calls (\eg POST requests via XMLHttpRequest) or those based on Chrome API endpoints. Leakage based on HTTP referrer header is assumed as accidental and is not considered as spying behavior. 

Table~\ref{tab:spyingextdef} gives examples of 5 spying extensions with an apparent disconnect between the claimed functionality of each extension and the user information they collect. For example, an extension called ``Pacman"\footnote{\url{chrome.google.com/webstore/detail/pacman/ffhefikmbepljajkbhedocnmgagdpajo}} provides a launch icon for a game. However, this extension spies on the browsing history of the user and does not provide any description about requiring browsing history for its functionality. We observe that all spying extensions in our dataset have a similar trait, thus suggesting leakage of user information with potential malicious intent.

\begin{table*}[]
\centering
\footnotesize
\begin{tabular}{p{2.25cm}llp{7.5cm}p{2.5cm}}
\toprule
Extension                          & Developer         & \# Users                   & Claimed Functionality                                                                                                   & Information Stolen                                                                                                               \\ \midrule
Block Site & wips & 1,072,111    & ``automatically blocks websites of your choice" & Browsing History \\
HolaSoyGerman              		   & topapps           & 75,962    & ``... know when a new German video rises, either from HolSoyGerman [YouTube channel]..."                    & Browsing History                                                                                    \\
SwytShop                           & swtyshop          & 35,015                      & ``automatically finds lower prices while you shop"                                                                         & Browsing History \\
Pacman                             & wips              & 22,670 & ``one click on the icon and play it {[}Pacman{]} right in your browser"                               & Browsing History                                                                        \\
Koukis Youtube                     & oeurstudio.com    & 2,662  & shortcut to go to Kouki's YouTube channel                                  & Domains Visited                                                                                   \\
\bottomrule
\end{tabular}
\caption{Difference in claimed functionality and user information stolen by spying extensions. 
}
\label{tab:spyingextdef}
\end{table*}



\subsubsectitle{\textbf{Data Collection Pipeline:} } We first conduct a crawl of all available extensions on the Chrome store and download 43,521 extensions spanning 12 categories. We identify spying behavior by relying on human expert verification. Since this requires considerable manual effort, we first shortlist possible spying candidates among all the extensions in the Chrome store, and manually verify only the short-listed samples. We divide our data collection pipeline into three procedures. \ding{192}We first identify a candidate set of potential spying extensions by applying various heuristics. \ding{193} Next, the extensions in the candidate set are verified by three human experts by analyzing behavioral logs generated by running the extension in a controlled environment. \ding{194} We further expand the candidate using various signatures (\eg based on file names and source code) extracted from newly identified spying extensions. Steps \ding{193} and \ding{194} are repeated until no new spying extensions are found. 

\begin{figure}
\centering
\includegraphics[width=0.49\textwidth]{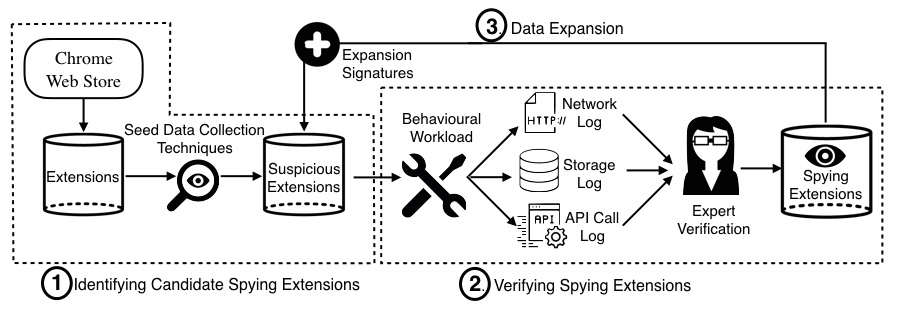}
\caption{Pipeline to identify spying extensions.}
\label{fig:datacollectionflow}
\end{figure}

\subsubsectitle{\textbf{\protect\ding{192} Identifying candidates for verification: }} 
We use the following 4 techniques to build a candidate set.
\begin{itemize}
\item \textit{Browser Extension Monetization Services:} There are extension monetization services that incentivize extension developers to insert Javascript-based user tracking code in the source code\footnote{\eg \url{http://ext.guru}, \url{https://monetizus.com}, \url{http://www.adonads.com}}. The tracking code is used by the monetization services to collect sensitive information about extension users, and developers are paid in exchange for such information. The monetization services may also inject advertisements on the client-side and developers earn a fraction of the generated ad revenue. Using web search with specific keywords (\eg \textit{browser extension monetization}), we collect a list of 15 popular monetization services. We short-list 115 extensions from the Chrome store that are developed by one of these 15 monetization services. In step \ding{193}, we verify 84 out of these 115 extensions to be spying. 

\item \textit{Filename Based Signatures:} We look for filename based cues to identify spying extensions. We use regular expressions like - *track*.js, *trk*.js, *user*.js, *stats*.js, *click*.js to locate files within the packaged code of extensions. We discover 79 extensions with suspicious file names matching our criteria and later find 3 to be spying. 

\item \textit{Permission Based Filtering:} Previous work has shown that certain permissions (and their combinations) listed in the Manifest file can be misused by developers to enable malicious behavior~\cite{kapravelos2014hulk}. We extract a candidate set of 150 suspicious extensions which ask for such permissions (e.g., background, webRequest, and activeTab). Among these extensions, we verify five extensions to be spying.



\item \textit{Reported Extensions:} In the past, there have been several media reports of spying extensions~\cite{Labs:2015dz}. We search the Chrome store for 30 extensions reported for spying in the last one year, and find 8 extensions to be still available on the store.
\end{itemize}


%

\subsubsectitle{\textbf{\protect\ding{193} Verifying whether an extension is spying: }}
After installing the extension, we apply an exhaustive web workload (covering the top 10 Alexa sites in English) that triggers a variety of browser actions (\eg form filling, tab open, mouse scroll) to collect extension behavioral logs. We collect information about network requests, changes to the client-side storage, and Chrome API calls to aid our manual human verification process. Chrome API calls are logged using the \textit{Chrome Extensions Developer Tool}.\footnote{\url{https://chrome.google.com/webstore/detail/ohmmkhmmmpcnpikjeljgnaoabkaalbgc/}} In order to capture networks requests generated by the extension, we use a \textit{record-replay} network tracing setup\footnote{\url{https://github.com/chromium/web-page-replay}}. In the \emph{record} phase, the workload is applied without loading the extension, and all network requests are cached. In the \emph{replay} phase, we run the workload with the extension loaded and serve requests from the cache when available and log any additional HTTP requests generated by the extension.



In our manual investigation process, we attempt to trace the flow of sensitive user information (\eg browsing history, form data) to third-party sites. We start by investigating the client-side storage changes and the Chrome API calls made by the extension to identify any access to user information. Next, we examine the network logs to determine any user information communicated to a third-party site as parameters of a GET request, or as payload in a POST request. One of the challenges here is that user information is sometimes communicated in an obfuscated form using various encoding schemes (\eg base64). We apply appropriate decoding schemes (whenever known) to identify such cases. We give more details of our verification technique in Appendix~\ref{verification}.

\subsubsectitle{\textbf{\protect\ding{194} Expanding the candidate set: }}
We extract various signatures from the spying extensions identified in step~\ding{193} to search for new extensions to be added to the candidate set. Signatures are extracted based on file names, developer identifiers, URLs in the source code, and Javascript code snippets. Next, we add any extension in the Chrome store with matching signatures to the candidate set. More details of our data expansion procedure are available in the Appendix \ref{datacol}.

Overall, we start with an initial candidate set of 374 extensions, and further expand it to 1,217 extensions after repeating steps~\ding{193} and~\ding{194}. In total, we identify 218 spying extensions.

\subsubsectitle{\textbf{Strengths and Limitations of Our Data Collection Methodology.}} The main strength of our approach is that we never flag an extension that does not steal user information as spying. However, given the manual effort required, our approach is not scalable. In addition, we may not be able to identify all the spying extensions in the Chrome store for two key reasons: \textit{First}, our candidate set building approach is based on a limited set of heuristics and is unlikely to shortlist all spying extensions on the Chrome store. \textit{Second}, spying extensions can use sophisticated encoding schemes to hide user information that we may not be able to identify.

 



To overcome these limitations, in Section~\ref{sec:detection}, we propose a machine learning based scheme that automatically detects spying extensions (without any manual effort). 

\subsection{Dataset Description} 
\noindent In this work, we use two types of datasets: (1) A near-complete dataset of all extensions available on the Chrome Web Store as of May 2016 (\texttt{CWS\_All}), (2) A dataset of spying Chrome extensions (\texttt{CWS\_Spy}). \texttt{CWS\_All} is collected by a crawl of the entire Chrome Web Store and \texttt{CWS\_Spy} is the dataset built using the methodology described previously, between the time period May-July, 2016. For both datasets, we collect the source code for each extension and perform an additional crawl to collect publicly available metadata associated with each extension. Metadata includes information about the size of the user base, reviews, ratings, bug reports, general functionality of the extension, and the developer. 

Table~\ref{tab:dataset-desc} shows high-level statistics for the collected data. The \texttt{CWS\_All} dataset contains 43,521 extensions (similar to count reported in prior study~\cite{kapravelos2014hulk}) spread over 12 primary categories. For our \texttt{CWS\_Spy} dataset, we identify 218 spying extensions spanning a diverse range of categories. It is interesting to note that spying extension developers target the top categories in \texttt{CWS\_All} (by volume of extensions), namely, ``Productivity'', ``Fun'', and ``Communication'', more than other categories, including ``News". The last column in Table~\ref{tab:dataset-desc} shows the name of the top spying extension in each category based on the number of current users. 

\begin{table*}[h]
\centering
\footnotesize
\begin{tabular}{lclll}
\toprule
Category        & \#Extensions (\% on Chrome Webstore)  &   Spying (\%) & Top Spying Extension			&	\#Users\\ 
				& \texttt{CWS\_All}&	\texttt{CWS\_Spy} &						&			\\ 	\midrule
Productivity    & 14,547 (33.42)    & 31 (14.22)     & Block site					&	1,072,111\\
Fun             & 6,613 (15.19)     & 80  (36.69)    & Channel Sub Box for YouTube	&	142,621 \\
Communication   & 5,764 (13.24)     & 35  (16.05)    & HolaSoyGerman           		&	106,022\\
Web Development & 4,162 (9.56)      & 5  (2.29)      & Web Developer Tools			&	2,081\\
Accessibility   & 4,062 (9.33)      & 7  (3.21)      & GouQi 						&	599 \\
Search Tools    & 2,493 (5.73)      & 6  (2.75)      & Handy maps 					&	6,451 \\
Shopping        & 2,060 (4.73)      & 7  (3.21)      & SwytShop						&	35,015\\
News            & 1,495 (3.43)      & 32 (14.68)     & Custom RSS news 				&	6,126\\
Blogging        & 835 (1.92)        & 11  (5.04)     & Koukis Youtube				&	2,662\\
Photos          & 651 (1.49)         & 1  (0.46)     & Take a picture everyday		&	200\\
Sports          & 569 (1.31)        & 3 (1.38)       & RotoGrinders Fan Duel Tool	&	7,818\\ 
By Google		& 62 (0.14)				& 0							&  &\\ \midrule
Total           & 43,521            & 218                 & & \\ \bottomrule
\end{tabular}
\caption{Distribution of spying and other extensions across various categories.} 
\label{tab:dataset-desc}
\end{table*}


\subsubsectitle{\textbf{Communication with Google.}} Considering the privacy risks facing users, we followed responsible disclosure procedure and established communication with Google. We shared a version of our draft and on 8 June, 2017 additionally shared the details of spying extensions identified in our work.


\section{Analysis of Spying Extensions} 
\label{sec:analysis}
\noindent We analyze various aspects of spying extensions, including its behavior, popularity and reputation, and developers. Our analysis serves as a first step towards understanding operational aspects of spying extensions before we explore techniques for automatic detection in Section~\ref{sec:detection}. 


\subsection{Modus Operandi of Spying Extensions}

\subsubsectitle{\textbf{Information Tracked.}}  \label{subsubsec:whatistracked}
\noindent Table~\ref{tab:infotrack} shows the different types of user information tracked by spying extensions. We find that most of the extensions steal browsing history of users (\ie complete URLs of sites visited), albeit there are two extensions that only track domains visited by the users. Information about browsing history can provide access to sensitive and private documents on services like Google Drive, Dropbox and Pastebin where a document can be accessed by anyone with a link to the document. We also found extensions stealing IP address, geolocation, and social media access tokens. In all cases, we found explicit code to steal such information. In fact, extensions were using social media access tokens to access private information of users, such as photos and posts with limited privacy settings.  

\begin{table}[h]
\centering
\begin{tabular}{@{}ll@{}}
\toprule
Information Tracked     & \#Total  \\ \midrule 
Browsing History        & 202 \\
IP Address, Geolocation & 10   \\
OSM Access Tokens       & 4   \\
Domain Visited          & 2   \\ \midrule 
Total Unique            & 218   \\ \bottomrule 
\end{tabular}
\caption{User information stolen by spying extensions.}
\label{tab:infotrack}
\end{table}

\subsubsectitle{\textbf{Spying Behavior.}} \label{subsec:capabilities} 
\noindent To understand how a spying extension works, we investigate the following aspects: (1) ability to access specific types of sensitive user information, (2) ability to store information (may not be used always), (3) ability to send tracked information to a remote server, (4) remote entities stealing the information, and (5) Remote Command-and-Control (CnC) setup used by spying extensions. 

In each part (whenever relevant), we analyze the specific permissions used by extensions to access various privileged Chrome API endpoints. To obtain this information, we first identify the different Chrome API endpoints required by spying extensions for accessing, storing and sending personal data (Section~\ref{sec:dataset}). We then map the different Chrome API endpoints used by an extension to specific permission requirements\footnote{Chrome Permissions~\url{developer.chrome.com/extensions/declare_permissions}}. 



\subsubsectitle{(1) Accessing sensitive user information.} Table~\ref{tab:access_cap} shows a list of permissions required for accessing sensitive user information. For comparison purposes, we also show the number (and percentage) of all other extensions ($\texttt{CWS\_All} \setminus \texttt{CWS\_Spy}$) that use the same permissions (possibly for other purposes). Spying extensions can continuously monitor user's browsing behavior with the help of `tabs', `activeTab' and `all urls' permissions which enable them to access the URL being visited at the moment by a user. We observe that the `tabs' permission ranks at the top (being used by over 94\% of spying extensions), but also used by a majority of all other extensions. Similarly, the `cookies' permission is also used by most (83\%) of the spying extensions (e.g., to access social media access tokens), but only by a smaller fraction (8.9\%) of all other extensions. Browsing history and location can also be tracked by direct access to `history' and `geolocation' permissions.

\begin{table}[h]
\centering
\begin{tabular}{llll}
\toprule
	 & 				 &\texttt{CWS\_Spy}    			&$\texttt{CWS\_All} \setminus \texttt{CWS\_Spy}$ \\
	 & Permission    & \#ext (\%)  			    	& \#ext (\%)    \\ \midrule
1    & tabs          & 207 (94.95) 					& 22,483 (52.24) \\
2    & cookies       & 181 (83.03)					& 3,844 (8.93)   \\
3    & storage       & 22 (10.09)  					& 13,376 (31.08) \\
4    & all urls      & 14 (6.42)   					& 4,022 (9.35)   \\
5    & history       & 6 (2.75)    					& 862 (2.0)     \\
6    & geolocation   & 3 (1.38)    					& 378 (0.88)    \\
7    & activeTab     & 3 (1.38)   					& 5,920 (13.76)  \\
\bottomrule 
\end{tabular}
\caption{Permissions to access user information.}
\label{tab:access_cap} 
\end{table}

\subsubsectitle{(2) Storing sensitive user information.} 
Spying extensions may or may not store user information on the client side. If they do, client-side cookies and `unlimited storage' permissions are used to store information either in plain text or in obfuscated form. This data is stored and accessed by extensions at regular intervals or periodically by a CnC remote server controlling the extension. We find that most of the spying extensions (188 out of 218) are storing user information before sending it to remote servers. We observe that 83\% of spying extensions access `cookie' permission (as compared to only 8.93\% other extensions) and 7\% of spying extensions have access to `unlimitedStorage' (compared to 6\% other extensions). Thus, inspecting local storage could be useful to identify spying behavior.  

\subsubsectitle{(3) Sending sensitive user information.} Spying extensions use various techniques to send user information to third party sites. We find 64\% of spying extensions using Chrome's WebRequest permission to send network request as a POST query with an obfuscated payload containing user information. The remaining extensions use an XMLHttpRequest Web API call, instead of a Chrome API call. 

\subsubsectitle{(4) Remote spying entity.} We find that the 218 spying extensions send sensitive user information to 29 unique (TLD+1) domains. While most of the extensions contact a single domain, there are six extensions contacting more than one domain, thus sharing user data with multiple entities. 
The top 4 domains in our dataset are \url{wips.com}, \url{upalytics.com}, \url{analyticssgoogle.com} (different from Google Analytics) and \url{fairsharelabs.com}. 


To understand the reputation of the tracking domains, we query three popular blacklisting services---Google SafeBrowsing API~\cite{Google:ij}, VirusTotal~\cite{Virustotal:kk} and Phishtank~\cite{Phishtank:pi}. We also analyze their WoT (Web of Trust~\cite{wot}) ranking which is a crowd-sourced website reputation metric. None of the domains are blacklisted by any of the services, indicating that spying extensions are difficult to identify based on their tracker domains. In fact, eight tracker domains have a WOT score of over 9.0 (out of 10.0) which indicates very positive crowdsourced feedback.

\subsubsectitle{(5) Control of spying extension by a CnC server.} We observe that some extensions do not start spying on users at the time of installation. They need to be triggered by a remote server to either start collecting user data or to send the stored user data to the remote server. We observe two methods being used to control spying state: (1) by toggling flags in cookies and client-side storage, and (2) time-based triggers which sends information after a certain period of time. This functionality makes detection of spying extensions a challenging task as they may act benign during some period of time. We identify 88 extensions where the tracking code was toggled on and off using a state variable in the cookie store via the `cookie.set' API call. During communication, the remote server can change the cookie state to control spying behavior. We find 2 extensions using time-based triggers which start tracking after fixed periods according to a time defined in a source JavaScript file. These control triggers require the `webRequest' and `cookies' permissions. The remaining 128 extensions did not require any specific triggers to start spying.

\subsection{Popularity and Reputation}
\noindent We analyze the popularity and reputation (based on crowdsourced feedback) of spying extensions and compare them with all extensions on the Chrome Store.

\begin{figure}
\centering
\includegraphics[width=0.48\textwidth]{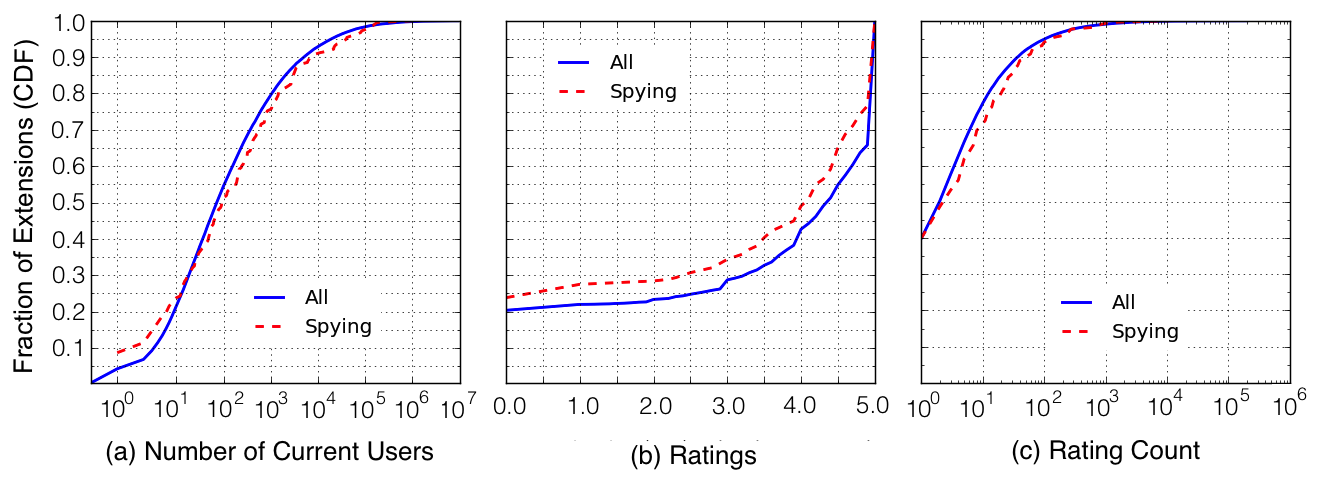}
\caption{(a) Number of current users for all the Chrome store extensions, (b) Crowdsourced average rating of extensions on Chrome Store, (c) Number of votes (count) contributing rating of extensions on Chrome Store.}
\label{fig:metainfoanalysis}
\end{figure}

\subsubsectitle{\textbf{User Base and Ratings.}} Spying extensions account for over 2.4 million cumulative installs, with a median rating of 4.4. Figure~\ref{fig:metainfoanalysis} shows the distribution of various popularity statistics. Spying extensions have a similar distribution of number of users, average rating, and number of ratings as all other extensions on the Chrome store. Thus, it would be hard to distinguish between spying and benign extensions based on popularity metrics by both Google (browser platform provider) and by users deciding whether to install an extension. Spying extensions include 190 extensions with over 10k users, and 1 extension with over a million users. This is alarming considering the privacy risk. Recently, Web-of-Trust extension with 140 million users was taken down for spying~\cite{Lifehacker:pi}. 

\subsubsectitle{\textbf{Bug Reports, Questions and Complaints.}}
Chrome Web Store allows users to submit questions or post suggestions and problems pertaining to extensions. We crawl these text reports associated with each extension and analyze it to identify if users are reporting bugs or complaining about the functionality of the extension. Out of 218 extensions, 103 (47\%) extensions received at least one comment by a user. We scan the text reports using an extensive set of keywords (based on manual investigation) indicating suspicious behavior (e.g., `tracking', `keylogger', `steal', `fake', `don't install', `malware'). Interestingly, only 12 (5.5\%) extensions received some form of comment reporting suspicious behavior. As spying behavior is hidden in background browser activities and traffic, it would be difficult for non-expert users to identify and report such behavior. Our findings suggest that users are mostly unaware of spying behavior. 

\subsection{Developers} 
\noindent There are 69 unique developers for our dataset of 218 spying extensions. While a large fraction of developers (89\%), develop only a single extension, we were surprised to find that the remaining 11\% of developers account for a majority (65\%) of extensions. In fact, this small fraction of developers rank among the top 100 developers on the Chrome store based on total number of extensions. Hence, even popular developers can have a few spying extensions that go undetected and extensions from popular developers can not be trusted to be safe. We provide more details of a few spying extensions by popular developers in Appendix~\ref{topdev}. Recently, a few extensions by highly rated developers, wips.com and hoverzoom.com were taken down from the Chrome store for spying~\cite{Labs:2015dz}.


\subsection{Key Takeaways}
\noindent We highlight the key takeaways from our analysis that would help us build a robust defense (Section~\ref{subsec:rnn}).
\begin{itemize}
\item We identify the permissions that are misused to access, store and transmit user information. Later in Section~\ref{sec:detection}, we use these findings to identify a dataset of benign extensions (non-spying) and develop a machine learning based model to detect spying behavior.
\item Extensions can switch between spying and non-spying states based on remote and time based triggers. We devise our detection scheme to capture such dynamic spying behavior.
\item Popularity and other crowd-sourced reputation metrics serve as unreliable signals (prone to manipulation) for differentiating spying extensions from others. Thus, we do not use these metrics to build a detection scheme.
\end{itemize}

\section{Automatically Detecting Spying Extensions} 
\label{sec:detection}
\noindent In this section, we explore the use of supervised machine learning to automatically detect spying extensions. We investigate effectiveness of prior approaches, before proposing a new detection scheme that performs better. 

\subsection{Training and Evaluation Setup}
\noindent We begin by building a dataset of spying and benign (non-spying) extensions which can be used to train and evaluate ML classifiers. Below, we discuss the positive (spying) and negative (benign) samples used in the ML pipeline.

\subsubsectitle{Positive class. } We use previously identified 207 spying extensions (in Section~\ref{sec:dataset}).\footnote{Remaining 11 extensions (out of 218) were taken down at the time of this analysis, and thus could not be included.}

\subsubsectitle{Negative class. } We use observations from Section~\ref{sec:analysis} to identify benign extensions. \textit{First}, we include 10,131 extensions that do not request any of the permissions misused by spying extensions to access, store and send user information (Table~\ref{tab:infotrack}). \textit{Second,} we include 11,235 extensions that do not make any network requests. These extensions might access user information, or perform other browser operations, but do not make network requests. \textit{Lastly,} we add 831 extensions that were marked as not spying in our manual investigation (Section~\ref{sec:dataset}). Note that these 831 extensions add diversity to the benign class because they were initially shortlisted as suspicious, but verified as benign on manual investigation. These extensions are actually very similar to spying extensions and most of them make network requests and/or also access private information. Therefore, including them makes the detection task more challenging, and helps to build a better classifier. In total, we identify 16,785 unique benign extensions. We acknowledge that the representativeness of our negative class is potentially limited by the usage of heuristics to select extensions. However, classification performance would only increase if we have a more diverse negative class.

To reduce the high class imbalance, we use a 1:5 ratio of positive and negative class and repeat our training/validation experiments with different samples of benign extensions. We randomly choose three non-overlapping samples from the benign dataset, each of size 1,035 extensions. We report average precision (percentage of detected extensions that are spying) and recall (percentage of spying extensions that are detected) over the three runs of the experiment, based on 5-fold cross validation.


\subsection{Using Prior Work} \label{subsec:prior}
\noindent Prior work by Jagpal \textit{et al} used machine learning to automatically detect malicious extensions---a broader class of harmful extensions including those that engage in ad manipulation and social network abuse~\cite{jagpal2015trends}. In this section, we first assess the performance of Jagpal \textit{et al}'s approach on our dataset of spying extensions. For a fair comparison, we use the same set of features, classifiers and parameters used by Jagpal \textit{et al}'s work, as mentioned below.


\subsubsectitle{Feature Engineering. }Prior work by Jagpal \textit{et al.}~\cite{jagpal2015trends} takes significant effort to carefully select and model an exhaustive set of features. We begin by extracting a similar set of features to detect spying extensions. All the features are described in detail in Table~\ref{tab:feat-desc}. Note that our feature set includes all the features (whenever available)\footnote{For example, WoT reputation score of remote URL will be absent for extensions which do not make a query to an external URL.} used by Jagpal \textit{et al}~\cite{jagpal2015trends}. Features are broadly divided into two categories---static and dynamic.
\begin{enumerate}[leftmargin=*]
\item \textit{Static Features: } These features include permissions listed in the manifest file (F1), signatures extracted from the source code (F3) and meta-information like the number of users, reviews, ratings, etc. (F6). Static features provide a rich insight into the functionality and structure of extensions. 

\item \textit{Dynamic Features: } These features capture the traits of an extension when it is actively running on the client-side. Every extension is run in a controlled environment (using the same workload used in Section~\ref{sec:dataset}) to record client-side changes (F4), network behavior (F5), and the Chrome API calls made by the extension (F2).
\end{enumerate}

\begin{table*}[t]
\centering
\begin{tabular}{@{}lll@{}}
\toprule
                    & Feature Set                          & Feature Description                                                                                                                  \\ \midrule
\multirow{1}{*}{F1} & \multirow{1}{*}{Permission}          & \textit{Permissions:} One-hot vector for manifest permissions \\ \midrule

\multirow{1}{*}{F2} & \multirow{1}{*}{Chrome API Calls}    & \textit{API Call:} One-hot vector for invoked Chrome API calls    \\ \midrule
\multirow{2}{*}{F3} & \multirow{2}{*}{JavaScript Based}    & \textit{Eval:} Boolean val. for presence of eval function used in source code                                            \\
                    &                                      & \textit{base64:} Boolean val. for presence of base64 obfuscation                                                 \\ \midrule
\multirow{4}{*}{F4} & \multirow{4}{*}{Client Side Storage} & \textit{Cookies:} Number of cookie values changes                                                   \\
                    &                                      & \textit{Storage:} Number of localstorage changes                                     \\
                    &                                      & \textit{URL in Cookies:} Boolean val. for presence of URL in cookies                                                                  \\
                    &                                      & \textit{URL in Storage:} Boolean val. for presence of URL in localstorage                                                  \\ \midrule
\multirow{5}{*}{F5} & \multirow{5}{*}{Network Log}         & \textit{XML HTTP:} Number of XHR calls made by the extension                                                                                  \\
                    &                                      & \textit{GET:} Number of GET queries invoked by the extension                                                                                  \\
                    &                                      & \textit{POST:} Number of POST queries by the extension                                                                                        \\
                    &                                      & \textit{WoT:} Crowd-sourced reputation of remote URL\\ \midrule 
\multirow{2}{*}{F6} & \multirow{2}{*}{Others}              & \textit{Filename Match:} One-hot vector for suspicious filenames                       \\
                    &                                      & \textit{Metadata:} Rating, number of reviews, number of users                                                              \\ \bottomrule 
\end{tabular}
\caption{Features used for automated detection of spying extensions.} 
\label{tab:feat-desc}
\end{table*}

\begin{table*}
\centering
\begin{tabular}{l|l|c|l|c|l|c|l|c|l|c|l|c}
\hline
\multirow{2}{*}{Features} & \multicolumn{2}{p{1.8cm}}{\textbf{LR (used in prior work)}} & \multicolumn{2}{|p{1.8cm}}{DecisionTree} & \multicolumn{2}{|p{2.3cm}}{Random Forest} & \multicolumn{2}{|p{1.9cm}}{SVM (rbf)} & \multicolumn{2}{|p{1.9cm}}{Adaboost} & \multicolumn{2}{|p{2.3cm}}{Neural Network} \\ \cline{2-13} 
                          & P        & R	& P        & R        & P         & R        & P       & R      & P      & R      & P         & R        \\ \cline{1-13}
F1                        & 9.10 & 12.43 & 9.17             & 10.18           & 13.27              & 16.02         & 13.92            & 15.68       & 14.87          & 16.21       & 65.32             & 69.87          \\
F2                        & 18.12 & 16.23 & 17.92            & 13.22         & 24.89             & 25.13         & 25.27           & 25.83       & 26.21          & 24.93       & \textbf{72.04}             & \textbf{75.20}         \\
F3                        & 18.00 & 16.01 & 17.35            & 16.72         & 22.16             & 21.12         & 16.22           & 35.11       & 17.16          & 31.81       & 64.81             & 61.23         \\
F4                        & 15.28 & 18.15 & 15.15            & 17.13         & 15.06             & 19.02         & 17.32           & 23.92       & 20.82          & 32.31       & 64.19             & 61.11         \\
F5                        & 15.09 & 14.51 & 18.09            & 14.08         & 22.8             & 20.15         & 15.12           & 31.99       & 24.83          & 25.15       & 67.09             & 64.28         \\
F6                        & 18.18 & 19.44 & 20.35            & 19.81         & 22.18             & 19.92         & 14.27            & 15.81       & 22.30          & 17.17       & 65.21             & 64.12         \\
All Feats                 & \textbf{22.35} & \textbf{24.18} & 21.73            & 20.19         & 23.96             & 23.99         & 24.83           & 26.79       & 30.03          & 32.32       & \textbf{78.12}             & \textbf{80.32}         \\ \hline
\end{tabular}
\caption{{P(=Precision) and R(=Recall) in \% for detecting spying extensions using machine learning classification.}}
\label{tab:ml}
\end{table*}

\subsubsectitle{ML Classifiers. } In addition to using the classifier used by Jagpal \textit{et al.}, which is Logistic Regression (LR) with L1 regularization, we experiment with a variety of other high performing classifier families. We include Decision Trees, Random Forest, Adaboost (an ensemble method using Random Forest as the base estimator), SVM (with an RBF kernel) and a Neural Network. For the Neural Network, we use an architecture with a single hidden layer having 100 neurons and use stochastic gradient descent with a reLU activation function. We also use L2 regularization with a high penalty value, $\alpha = 1.0$ to limit overfitting.

\subsubsectitle{Detection Performance. }
Precision and recall values for the spying class are reported in Table~\ref{tab:ml}. LR-based classifier used by Japgpal \textit{et al.} performs poorly over the entire feature set yielding very low precision and recall of 22.35\% and 24.18\%. Thus this technique based on prior work is ineffective against spying extensions.

Except Neural Network, all other classifiers also yield low detection performance. Neural Network using all features provides the highest precision of 78.12\%, and a recall of 80.32\%. More importantly, among the different feature categories, we obtain the highest classification performance with category F2, which are features based on Chrome API calls. Thus there is scope for improving detection performance, if we can better leverage the Chrome API call feature category. We investigate this further in the next section.






\subsection{Our Idea: Leverage Sequential Patterns in API Calls}

\subsubsectitle{Key Insight.} Our key insight is that we can distinguish spying extensions from other extensions based on patterns in Chrome API call sequences. 

\begin{figure}[h]
\includegraphics[width=0.45\textwidth]{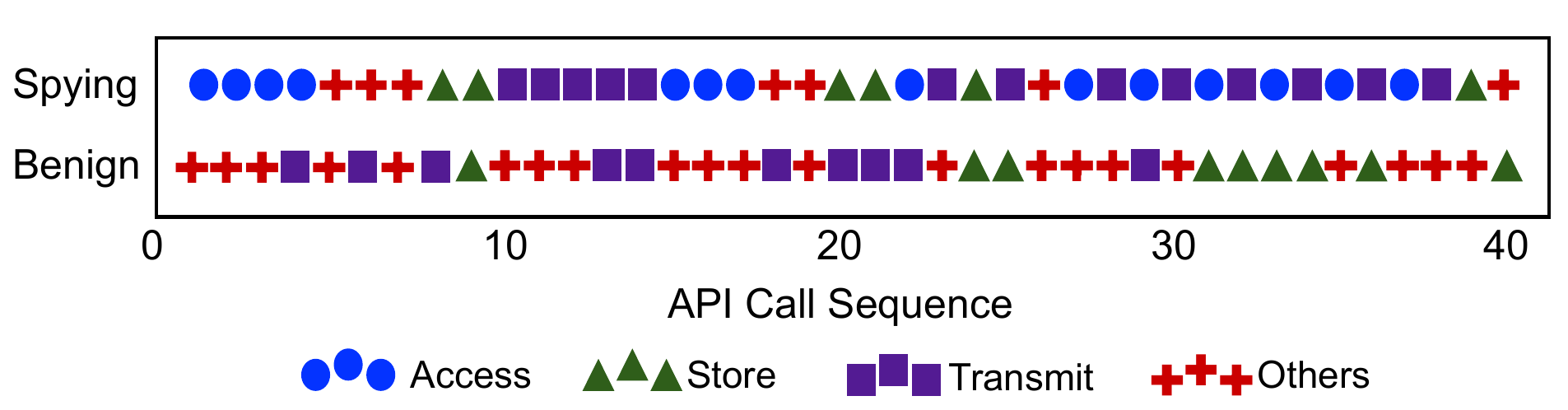}
\caption{Time series of events for a sample of spying and benign extensions.}
\label{fig:timeseries_visualization}
\end{figure}

We provide a simple illustration of our insight using a spying and benign extension from our dataset. Figure~\ref{fig:timeseries_visualization} shows the Chrome API call sequences for these extensions. For better understanding, the API methods are color coded based on whether they are \textit{access}, \textit{store} or \textit{transmit} methods. Access methods are those associated with the following API endpoints --  bookmarks, cookies, history, storage, and tabs. We focus on these specific endpoints because our dataset of spying extensions uses these endpoints for accessing user information. Store operations includes the following persistent storage endpoints--- bookmarks, cookies, history and storage. Transmit methods include those used to send and receive HTTP requests. The remaining API methods are annotated as \textit{others}. The sequence of these color coded events is shown across the X-axis, denoting the API call sequence events of the specific extension.

In Figure~\ref{fig:timeseries_visualization}, there is a striking difference in the patterns of spying and benign extensions. The spying extension has access operations closely followed by a transmit operation. The benign extension on the other hand exhibits network transmit operations like spying extensions but does not access any user information.

Note that the above is a very simplistic example of sequential patterns exhibited by spying and benign extensions. In practice, these patterns could be more complicated. For example, access and transmit methods may be further apart in the case of spying extensions, or there might be benign extensions that also use access and transmit methods in different ways. In fact, we even do not want to restrict ourselves by annotating API methods as access, store and transmit methods. \textit{Thus, the key challenge is to automatically distinguish between spying and benign extensions based on any available complex patterns in API call sequences.}

\subsubsectitle{Learning Complex Patterns in API Call Sequences to build a Classifier.} Previous study by Canali \textit{et al.} surveyed use of n-gram and bags model on system API calls to extract signatures from malware for malware detection. They found that N-gram models reach a performance threshold after a certain sequence length and triggers more false positives as the sequence length increases~\cite{Canali:2012:QSA:2338965.2336768}. To verify if n-grams could yield significantly better detection performance, we repeat our experiments in Section~\ref{subsec:prior} by including n-gram of API calls made by extensions (of size 2 and 3) as a feature. We observe only a marginal increase of 1.01\% precision and 1.23\% recall when using the Neural Network based model with all the existing features along with n-grams. Therefore, we need a model that can better identify complex sequential patterns.

We leverage recent advances in deep neural networks, and use a \textit{Recurrent Neural Network (RNN)} to learn complex patterns in sequential data. In fact, RNN models outperform traditional n-gram based models when modeling sequential data~\cite{de2015survey}. In our case, an RNN based classifier takes in as input, a sequence of API call names to distinguish between spying and benign extensions. Note that the input sequence does not include any other metadata associated with API calls, such as timing information or call parameters. 




\subsubsectitle{Benefits of using API Call Sequence. } We further motivate the usefulness of API call sequences (and an RNN) by highlighting three key advantages.

\begin{itemize}[leftmargin=*]
\item \textit{Abandon need for hand-crafted features. } Learning patterns in API call sequences using an RNN obviates the need for extensive feature engineering as in previous research. We show that an RNN-based approach can achieve high detection performance using only sequential data of API calls.
\item \textit{Call sequence captures core spying behavior. } Chrome API calls are required for the core functionality and operation of an extension and especially for enabling spying capabilities. Therefore, it would be difficult for a developer to adapt the API call sequence to bypass a defense (Section~\ref{subsec:rnn} analyzes robustness of defense). On the other hand, defenses based on static features would not be so robust. For example, we have seen in Section~\ref{sec:analysis} that crowd-sourced reviews and ratings of spying extensions are similar to that of benign extensions. Even other static features like file name based signatures can be easily adapted by a developer to bypass defenses. 
\item \textit{Adapt to dynamic spying behavior. }A spying extension may not be triggered at the time of evaluation. Chrome API calls can capture changes in the behavior of an extension which may periodically switch between spying and benign activities. Thus a defense based on API call sequence can adapt to dynamic spying behavior. 
\item \textit{Chrome API call sequence encompasses most other dynamic features. }
Though there exist other dynamic features apart from Chrome API call sequences viz. network log, change in and client-side storage; API call based features form a bigger umbrella encompassing other dynamic features. Most of the operations captured by network logs and cookie/storage logs also reflects in the Chrome API call logs with a corresponding entry, with the exception of Web API calls (which is available to all web pages). An example of a Web API call is an XMLHttpRequest to send and receive information from a remote server. Later, in Section~\ref{subsec:rnn}, we show how we can augment Chrome API calls sequences with Web API calls to further improve detection performance. By tracking such network requests, note that we can also potentially detect spying behavior driven by content scripts.
\end{itemize}

\subsection{RNN based Detection Scheme} \label{subsec:rnn}

\noindent We present a Recurrent Neural Network (RNN) based sequence classification approach to accurately detect spying extensions. In this section, we explain our proposed model, assess its performance, and compare it with existing techniques. 

\subsubsectitle{RNN Background. } Unlike traditional Neural Networks, a Recurrent Neural Network (RNN) has a notion of ``memory'' that captures information seen so far. An RNN has feedback connections which allow the model to have loops connecting the output layer to the input, enabling the model to build a memory about input fed to the network over time. This enables RNN models to learn patterns in sequential data such as natural language text or time-series data. 

A basic RNN takes a sequence as input, and then at each training step updates a hidden state (that captures the memory) to generate an output vector that can be used to predict the next item in the sequence. Each hidden state is iteratively updated based on the previous hidden state and the current input so that the output vector better predicts the next item in the sequence. While the basic RNN is efficient at capturing short-term sequential patterns, it is unable to capture long-term dependencies (or patterns). We use a special variant of RNN, called Long Short Term Memory networks (LSTM) that is capable of capturing long-term dependencies~\cite{hochreiter1997long}.

An LSTM has multiple memory cells, where each memory cell has special input and forget gates that can capture long-term dependencies. Using the forget and input gate, the hidden state of each memory cell is iteratively updated to forget a fraction of existing memory and to add new memory, respectively, to better predict the next item in a sequence. LSTMs have been shown to outperform traditional sequence classification techniques such as Hidden Markov Models (HMM)~\cite{hochreiter1997long}. LSTM has also been shown to effectively capture dynamic sequential behavior to detect malicious applications~\cite{pascanu2015malware}\cite{shibahara2016efficient}. 

\subsubsectitle{RNN Model Architecture. } We use an LSTM model for our detection task. As input to the LSTM model, we use the sequence of Chrome API calls made to specific API endpoints at a method level. The only pre-processing we perform on the input sequence is converting each API call (e.g. \emph{cookie.set}) to a unique numerical identifier. We use two hidden LSTM layers with 200 hidden units in each layer, with a dropout rate of 0.2 (to avoid overfitting) and a reLU activation function at the output layer. Learning rate is tuned using RMSprop gradient descent optimization and training loss is computed using the cross entropy loss function. 

\begin{figure}[h]
\centering
\includegraphics[width=0.4\textwidth]{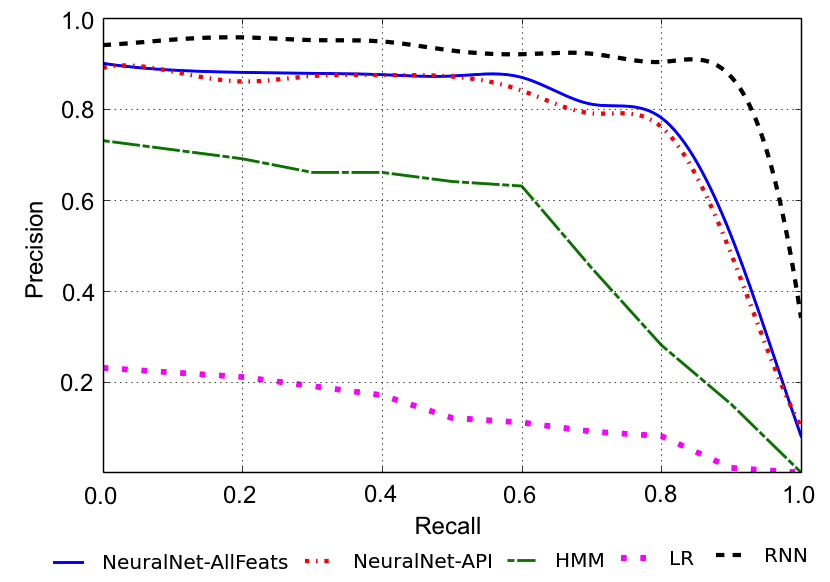}
\caption{Precision-Recall curve using API call sequences as a feature to detect spying extensions.}
\label{fig:lstm-results}
\end{figure}

\subsubsectitle{RNN Detection Performance. } Table~\ref{tab:model-comarison} shows comparative results and Figure~\ref{fig:lstm-results} shows corresponding Precision-Recall curves. For comparison, we present results for a HMM model using API call sequence as input, and Neural Network and Logistic Regression based models using all features (in Table~\ref{tab:model-comarison}). Our RNN based approach outperforms other approaches with the highest precision (90.02$\pm$1.06\%) and recall (93.31$\pm$0.92\%).
From the precision-recall curve, it is also clear that we need not significantly sacrifice precision to improve recall for our approach. 

\begin{table}
\centering
\begin{tabular}{@{}lll@{}}
  \toprule
               & Precision (\%) & Recall (\%)\\ \midrule
  Neural Network-All      & 78.12     & 80.32  \\
  Neural Network-API      & 72.04     & 75.20  \\
  HMM          & 66.23     & 63.27  \\
  Logistic Regression (LR)	& 22.35	& 24.18 \\	
  RNN    & \textbf{90.02}     & \textbf{93.31}  \\ \bottomrule
  \end{tabular}
\caption{\small{Precision and recall comparison of various models.}}\label{tab:model-comarison}
\end{table}

\subsubsectitle{Analysis of False Positives. } 
There are 22 benign extensions marked as spying by our model. Among them, 17 extensions access and transmit user information to a remote server. However, when we corroborate this behavior with their description on the Chrome Web Store, we find that such access is required for the functionality of these extensions. For example, one such extension is called Honey\footnote{\url{https://chrome.google.com/webstore/detail/bmnlcjabgnpnenekpadlanbbkooimhnj}}, which is a popular extension to find discount coupons on e-commerce websites. This extension requires to capture the URLs visited by the user for its operation. We did not find any conclusive reason for the remaining 5 benign extensions marked as spying. 

\subsubsectitle{Analysis of False Negatives. } There are 7 spying extensions that we are not able to detect. Among them, 2 extensions generated a very small trace of API calls (less than 55 API calls, compared to the average call length of 215). Possibly, these sequences were not long enough for the RNN model to effectively identify patterns. For these cases, running a longer workload could trigger more API calls, and potentially lead to detection. Next, there are 5 spying extensions that do not use Chrome API calls to transmit user information, but instead use a Web API call (XMLHttpRequest) to make network requests. However, it is easy to detect these extensions by augmenting the Chrome API call sequence with Web API calls (discussed next).

\subsubsectitle{Augmenting Chrome API calls Sequences with Web API Calls. } One limitation of our approach (using Chrome API call sequences) is that we do not consider the network requests made using the Web API, mainly the XMLHTTP Requests (XHRs) to send user information. To address this limitation, we perform an additional experiment where we train an RNN by merging XHR calls and Chrome API calls (by ordering the calls based on timestamps). Using this model, we achieve a higher recall of 95.02\% and a higher precision of 91.32\%, compared to the RNN model trained only using Chrome API calls. Hence, our approach can be easily extended to further improve detection performance by incorporating other Web API calls into the RNN input sequence.

\subsubsectitle{Robustness of RNN Model. }An attacker can try to evade detection by perturbing the API call sequence. To test the robustness of our RNN model against such attacks, we perturb the API call sequences of spying extensions and evaluate impact on detection performance. Perturbation is done by inserting a `benign' sequence of certain length at a random position between the first instance of information access and first instance of information transmit. In practice, an attacker strategy might be similar where he tries to hide data leakage by generating noise sequences before sending out the information to a third party. The sequence to be inserted is a randomly chosen n-gram sequence from non-spying extensions. We repeat the experiment for different lengths of inserted sequence ranging from 1 to 200 (215 is the average number of API calls for all extensions using our workload), and perform 5 random trials for each spying extension. We observe that detection performance drops slightly as we increase the length of the inserted sequence. For the longest inserted sequence of 200, our precision drops to 86.08\% (from 90.02\%) and recall drops to 89.9\% (from 93.31\%). This indicates that the RNN model is quite robust against heavy perturbations. 

Theoretically, an attacker can further increase the perturbations to evade detection. However, the downside of such a heavy perturbation would be reduced performance of the browsing session, which might alert the user. The browser can also impose rate limits on the API endpoints to limit the number of operations done by an extension, to limit abuse and avoid browser slowdown. In addition, if the attacker increases the perturbations heavily, the extension might become easier to detect by existing machine learning models, as the activity patterns might look abnormal. An attacker can also try to evade detection by reducing the number of API calls. However, this would only force the spying extension to behave normally, and potentially limit information leakage.

\subsection{In-the-Wild Detection of Spying Extensions} \label{subsec:inwilddetection}
\noindent To discover new spying extensions from rest of the extensions on Chrome Web Store, we use our trained model to conduct an in-the-wild detection of spying extensions. We apply our RNN classifier to 42,110 extensions on Chrome-Store. Our RNN model marks 90 extensions as spying. A manual investigation and verification confirms 65 extensions to be spying. Here we give details about these 90 extensions marked as spying --

\subsubsectitle{Extensions confirmed to be spying: } Manual verification confirms 65 out of 90 extensions to be actually spying. We found all the spying extensions sending browsing history to external URLs, when such an operation was not needed for their functionality. These 65 spying extensions had a user base of 4.2 million and 4 of them were taken down after our analysis. Table~\ref{tab:falsepos} gives details of a sample of these newly found spying extensions which are most popular.

We suspect that there are more spying extensions in the Chrome Store. We could potentially identify more extensions by increasing the size of our ground-truth dataset (currently only 207 spying extensions). Neural networks become more effective with a larger training dataset. Unlike Google, we do not have resources to build ground-truth at scale based on our manual scheme. In fact, prior work by Google (Jagpal \textit{et al}) showed that their recall increased from 77\% to 91\% by adding newly discovered malicious extensions (identified over a 2 week period) into the training set~\cite{jagpal2015trends}. 



\subsubsectitle{Unverified: } Among the remaining 25 extensions, we were unable to confirm spying or benign behavior for 5 extensions. The data transmitted by these 5 extensions was highly obfuscated, and we were unable to decode the network requests.

\subsubsectitle{Benign but marked as spying: } We verify the remaining 20 extensions to be benign. These extensions were accessing and sending user information to remote servers, but such an operation was needed for the functionality of the extension. It is also important to note that our False Positive Rate is extremely low since we only find 20 false positives after testing on 42k extensions.


Overall, our in-the-wild experiment shows that we are able to catch several new spying extensions using a limited training dataset. The operator can further boost the performance of the model by adding new training samples from newly found spying extensions.

\newcommand{\para}[1]{\vskip0.075in\noindent {\bf{#1}}}

\section{Discussion: Defense at the Edge}
\label{sec:edge}
\noindent Usually browser providers (like Google, Mozilla) detect malicious extensions by running the extensions in a controlled centralized environment~\cite{jagpal2015trends}, which is also possible with our detection scheme. In this section, we discuss the potential for an alternative approach of pushing the detection to the edge at the user's browser.

\subsubsectitle{Deployment Model. } We envision a deployment scenario where our pre-trained detection scheme is available on the user's browser as another extension.
In this model, the detection extension needs to be able to monitor API calls of other extensions. In the case of Chrome, this monitoring capability is only available to white-listed extensions (e.g., Chrome Apps \& Extensions Developer Tool). 
The detection tool can analyze API calls made by any extension on the user's browser (in real time) and alert the user if it detects spying behavior. 

\subsubsectitle{Benefits of Pushing Detection to the Edge. }  We discuss three key benefits.
\begin{itemize}[leftmargin=*]
\item \textit{Enable detection based on user generated workload.} Spying behavior may be triggered by certain types of usage patterns (e.g., stealing social media access tokens when user logs into a social media site). Client-side detection increases the possibility of covering these cases, making it harder for spying extensions to evade detection. 
\item \textit{Inform privacy leak to users. } In practice, users may have different expectations of privacy. Client-side detection allows users to be aware of the specific client-side data which is being accessed and transmitted by extensions, which can help them better understand privacy risks associated with the extensions. 
\item \textit{Early detection of spying behavior. } 
%
%
When spying extensions manage to evade centralized detection, client-side detection provides a second chance to detect them in the early stage. The question in this case is how quickly can a client-side detection method identify spying behavior, once a user starts using the extension. We design an experiment to understand the amount of user activity (or API calls) required to detect spying extensions. For each extension, we apply a workload that includes visiting 10 websites in sequence, and triggering on average 9 different browser events (\eg mouse movement, clicking, and form fill) per site. We observe that our approach can detect extensions with high precision (90.02\%) and recall (93.31\%), after analyzing only 28 browser events (or 189 API calls), which is just 20\% of the entire workload we used in the previous sections. This result illustrates the potential for early client-side detection. 

\end{itemize}

A client-side detection scheme would incur significant computational overhead if we run a full fledged RNN classifier on the browser and can impact a user's browsing experience. However, the browser provider can use ML model compression techniques to build a lightweight high performing ML scheme on the client-side~\cite{bucilua2006model}. Recent efforts have demonstrated the possibility of running deep neural network models on mobile processors, and achieving high performance while consuming very low power~\cite{jin2014efficient}. Researchers have also explored light-weight machine learning implementations for web browsers~\cite{meeds2015mlitb}. More recently, Google and Facebook released lightweight real-time deep learning frameworks for mobile platforms~\cite{TensorFlow:2017jf}\cite{Facebook:2016fz}. Therefore, a client-side defense looks promising, given the recent advances in optimized deep learning runtimes.

\section{Related Work}


\subsubsectitle{User Behavior Tracking.} There have been several studies of third-party user behavior tracking across the Web~\cite{englehardt2016online}\cite{lerner2016internet}\cite{mayer2012third}. Closely related to our work are recent studies by Starov \textit{et al.}~\cite{starovextended} and Weissbacher \textit{et al.}~\cite{weissbacher2017ex} that focus on browsing history leaking extensions. However, their study includes any extension that tracks browsing history, including benign cases. Instead, we conduct a detailed study of the more suspicious category of spying extensions, and also investigate other kinds of information leakage (OSN access tokens, geolocation). Starov \textit{et al.} do not propose a robust detection scheme for tracking extensions. Weissbacher \textit{et al.} detects browser history leaking extensions using a machine learning scheme based on n-grams of API call sequences. We propose a robust detection scheme based on Recurrent Neural Network (RNN) that can learn complex patterns in API call sequences.

\subsubsectitle{Detection of Malicious Browser Extensions.} While there are very few studies focusing on information leakage by browser extensions, researchers have studied a broader class of malicious extensions. These include extensions that exhibit rogue behavior, such as - injecting ads, spreading malware, OSN account hijacking, and information theft~\cite{kapravelos2014hulk}\cite{jagpal2015trends}\cite{xing2015understanding}. 

Prior work has also attempted to automatically detect malicious extensions. Kapravelos \textit{et al.} proposed use of HoneyPages to dynamically run extensions and trigger them to capture any malicious activity~\cite{kapravelos2014hulk}. Shahriar \textit{et al.} used an HMM model with code snippets from the source code of Firefox extensions as input to train a model for automatic detection of malicious extensions~\cite{shahriar2014effective}. However, this method fails in the case of extensions with obfuscated source code. Recently, DeKovan \textit{et al.} proposed a static heuristic based approach (using indicators from the source code of extensions) to flag malicious Chrome and Firefox extensions which change the layout of Facebook on client-side, or trigger suspicious user activity on Facebook~\cite{205855}. In addition, Jagpal \textit{et al.} proposed a machine learning framework with extensive feature engineering for malicious extension detection~\cite{jagpal2015trends}. In Section~\ref{subsec:prior}, we showed that this approach yields poor performance in detecting spying extensions. Moreover, their approach rely on complex feature engineering and rule based heuristics. Our RNN based approach captures dynamic behavior of extensions, and does not need hand-crafted features.

\subsubsectitle{Information Flow Control (IFC) for Malicious Behavior Detection.} There has been extensive research involving IFC techniques to detect security vulnerabilities and privacy leaks in JavaScript based applications \cite{jang2010empirical}. Information Flow Analysis for Javascript applications focus predominantly on taint analysis \cite{guarnieri2011saving}, unstructured control flows \cite{just2011information} and dynamic information flows \cite{hedin2012information}\cite{hedin2014jsflow}. Previous research has also explored IFC based techniques to discover security vulnerabilities in browser extensions~\cite{bandhakavi2011vetting} and detect malicious extensions which steal sensitive information, such as cookies and saved passwords~\cite{dhawan2009analyzing}. 
However, IFC based approaches face certain practical challenges. First, it is hard to find an exhaustive set of flow patterns that can effectively capture vulnerabilities. Second, they suffer from taint explosion and can result in a large number of false positives~\cite{enck2014taintdroid}. In our work, we take an alternate approach of leveraging the recent advances in deep neural networks to detect spying behavior.
\section{Conclusion}

\noindent This work presented a detailed study of browser extensions that steal sensitive user information. We started by building a dataset of 218 spying extensions on the Chrome store using expert manual investigation. Such a dataset allowed us to conduct an in-depth analysis of the modus operandi of these spying extensions.  
Being a privileged software residing in the user's browser, spying extensions can track online activities of a user. Spying extensions steal a variety of sensitive personal information, including the browsing history of the user. Surprisingly, these extensions are as popular as other extensions on the Chrome store. 
Our analysis thus helps to highlight the privacy risks faced by users of spying browser extensions. 

Next, we investigated techniques to automatically detect spying extensions using machine learning schemes. We found that an extensive set of hand-crafted features used in prior work for detecting malicious extensions are ineffective in our case. Instead, we showed that browser API call sequences serve as a robust feature to detect spying extensions. Using Recurrent Neural Networks that are best suited for learning patterns in sequential data, we showed that spying extensions can be detected with high precision and recall. Our RNN-based classifier identified 65 new spying extensions (which were not used for training the classifier) on the Chrome store. Lastly, we also discussed the potential for deploying our scheme directly on a user's browser for early detection of spying behavior.

\bibliographystyle{IEEEtran}
\bibliography{main}

\appendix \label{sec:appendix}

\subsubsectitle{(1) Spying Extension Data Expansion Techniques} \label{datacol}
After obtaining an initial set of 100 verified spying extensions, we apply the following data expansion techniques over these verified extensions to obtain a larger candidate set.
\begin{itemize}
\item \textbf{Filename Matching:} We extract the file names of the JavaScript files which are responsible for spying behavior. We search for extensions with similar file names (using substring matching) in the Chrome Web Store and find 384 more extensions out of which we verified 49 to be spying.
\item \textbf{Developer Based Signature:} We look for other extensions written by developers of the spying extensions in our seed dataset. We discover 159 such extensions, and verify 8 to be spying.
\item \textbf{URL Based Signatures:} From our seed dataset of spying extensions we create a list of remote URLs used to send over user information. We search for other extensions that include these URLs in the source code. Out of 145 matches, we verify 46 to be spying. For the remaining set, we suspect that these extensions were no longer tracking (or we could not trigger the spying mode). Also, some of the extensions were not functioning properly. 


\item \textbf{JavaScript Based Signatures:} From our seed dataset of spying extensions, we identify  JavaScript code snippets used for spying activities. We search for other extensions using similar code snippets. We find 155 closely matching extensions, and verify 15 to be spying.

\end{itemize}
\begin{table}[h!]
\centering
\footnotesize
\begin{tabular}{@{}lll@{}}
\toprule
\multicolumn{1}{c}{\multirow{2}{*}{Baseline}}       & \multicolumn{2}{c}{\#extensions} \\ \cmidrule(l){2-3} 
\multicolumn{1}{c}{}                                & Candidate       & Verified       \\ \midrule
permission based filter                             & 150             & 5              \\
by monetization services                            & 115             & 84             \\
reported extensions                                 & 30              & 8              \\
Filename based signatures                             & 79              & 3              \\ \midrule
Total (Baseline)                                    & 374             & 100            \\ \midrule
\multicolumn{1}{c}{\multirow{2}{*}{Data Expansion}} & \multicolumn{2}{c}{\#extensions} \\ \cmidrule(l){2-3} 
\multicolumn{1}{c}{}                                & Candidate       & Verified       \\ \midrule
filename matching                                   & 384             & 49             \\
more ext by developer                               & 159             & 8              \\
remote URL match                                    & 145             & 46             \\
JS based signatures                                 & 155             & 15             \\ \midrule
Total (Data Expansion)                              & 843             & 118            \\ \midrule
Total Spying                                        & 1,217           & 218            \\ \bottomrule
\end{tabular}
\caption{We identify spying extensions by first inspecting an initial dataset of extensions and then expanding the dataset using the expansion techniques.}
\label{tab:expansion}
\end{table}

\subsubsectitle{(2) Verification of Spying Extension} \label{verification}
Since it is not possible to manually investigate each of the 43k extensions on the Chrome Web Store, we use a combination of semi-automated heuristics and human verification to flag spying extensions. As the first step, we use a record-replay setup and automatically generate -- (1) network request log, (2) changes in client-side storage and (3) browser API calls made by the extension under review.

\balance
\subsubsectitle{\textbf{Capturing Network Request Log using Record-Replay.}}
\begin{itemize}
\item \textbf{Record Run:} In this phase, we prepare a behavioral suit for Top 10 Alexa websites. We use a clean session without any extensions, and devoid of any previous cookies, storage, cache or browsing history in this phase.

\item \textbf{Replay Run:} In the replay run we first load the extension in a new browser session and capture all the HTTP requests. The requests which generate 404 error help us to identify potentially untrusted remote URLs the extension is trying to contact. Next, we execute all the HTTP requests from the first step (live run). This enables us to capture the real behavior of extension when given Internet access. We log the resulting redirects and HTTP requests. 
\end{itemize}

\subsubsectitle{\textbf{Capturing Storage and Cookie States.}}
In the record run, we capture the local storage and cookie storage of the running browser session and compare it with the state after the replay run. This enables us to capture the storage and cookie changes caused by the extension under inspection. 


\subsubsectitle{\textbf{Capturing Browser API Access.}} 
To capture the browser API access by the extension, we modify the Chrome Apps and Extension Developer Tool which helps us to record details of all the API calls made by the extension along with the timestamps. 


\subsubsectitle{\textbf{Data Access, Storage, Sending Cues.}} 
\begin{itemize}
\item \textbf{Data Access:} We know the permissions and API calls which are needed to access user information (See Section~\ref{sec:analysis}). We look for those API calls from the log we generated to capture the API calls made by an extension. If any of those API calls are made then we mark the extension as trying to access user information.  

\item \textbf{Data Storage:} A spying extension may or may not store user data. We look for storage/cookie state before and after the extension is loaded and run to see if any data was stored by the extension on the client side.
\item \textbf{Data Transfer:}  If an extension accesses user information, we aim to check whether the `same' information is being sent to remote servers. As the first step, we check if any information is being sent to remote URLs by extracting the URLs and IP addresses from our network capture in the replay-run and live-run. Once we have these requests and the associated remote URLs, we try to decode the information that was sent to these remote URLs. We investigate the GET parameters and the POST request payload to identify data transfer. Sometimes, this information can be in plain text. If not, then we try to decode this information using decoding functions like base64, double base64 and hex. If the information being sent matches with the information being accessed, and the data access and transfer was happening beyond the functionality of the extension (as described on the Chrome Web Store), we mark the extension as `spying'.

\end{itemize}

\subsubsectitle{(3) Popular Developers and their Spying Extensions} \label{topdev}
Spying extensions by top eight developers (by number of users) have more than 1.5 million users. More details about these extensions, and the information stolen is given in Table~\ref{tab:topdevext}.

\begin{table*}[h!]
\centering
\footnotesize
\begin{tabular}{llp{2cm}lp{6cm}l}
\toprule
\multirow{2}{*}{Developer} & \multirow{2}{*}{\#SpyExt} & \multicolumn{4}{c}{Top Spying Extension}                                                                                                                                                                             \\ \cmidrule(l){3-6} 
                           &                           & Name                         & \#Users   & Description                                                                                                                                          & Information Stolen \\ \midrule
wips                       & 135                       & Block Site                   & 1,072,111 & ``automatically blocks websites of your choice"                                                                                                      & Browsing History   \\
muz.li                     & 1                         & Muzli 2                      & 107,655   & ``the freshest links about design and interactive, from around the web"                                                                              & Browsing History   \\
topapps                    & 5                         & HolaSoyGerman                & 106,022   & ``... know when a new German video rises, either from HolSoyGerman or JuegaGerman [YouTube channels]..." & Browsing History   \\
padlet.com                 & 1                         & Padlet Mini                  & 84,309    & ``collect and bookmark the best of web..."                                                                                                           & Browsing History   \\
wisestamp.com              & 1                         & Add Email Signature          & 68,651    & ``get a professional email signature.."                                                                                                              & Browsing History   \\
swytshop                   & 1                         & SwytShop                     & 35,015    & ``automatically,nds lower prices while you shop"                                                                                                     & Browsing History   \\
awesomescreenshot          & 1                         & Smart Shopper                & 32,625    & ``a universal shopping cart to allow cross-site shopping..."                                                                                         & IP Address         \\
rotogrinders               & 1                         & RotoGrinders - FanDuel Tools & 7,818          & Shortcut to manage rotogrinders.com games                                                                                                                                                     &  Browsing History                  \\ \bottomrule 
\end{tabular}
\caption{Top Developers (by number of users) and the top spying extension by each developer.}
\label{tab:topdevext}
\end{table*}


\subsubsectitle{(4) Generating workload for input to RNN} \label{subsubsec:workload}
For each extension, we pick a set of 4 websites (from Top 10 Alexa websites) and perform all browser actions in a controlled environment with the extension running in the background. Our workload also includes social media and email sites. We login to these social media and email sites using dummy accounts to increase chances of triggering any malicious activity by an extension. As a result of the dynamic run of the extension, we capture the network, and API call logs generated. We then extract the sequence of API calls from the API call log of each extension. To boost the data for training our classifiers, we repeat this experiment thrice for each extension with a different set of 4 websites each time, hence generating 3 set of sequences for each extension. As a result, we obtain 621 sequences (for 207 extensions) labeled as spying; and 9,315 sequences (for 3,105 extensions) labeled as benign. 

\subsubsectitle{(5) In-the-Wild Study - Newly Found Spying Extensions} \label{falsepos}
Our trained model when applied to the entire Chrome Web Store finds more spying extensions. Here we give details of top 10 (based on number of users) newly found spying extensions in Table~\ref{tab:falsepos}. 

\begin{table*}[h!]
\centering
\footnotesize
\begin{tabular}{p{3cm}llp{6cm}l}
\toprule
Extension                                 & Developer                  & \#Users & Claimed Functionality                                                                                                               & Information Stolen        \\ \midrule
Awesome Screenshot: Screen Video Recorder & awesomescreenshot          & 1,299,595    & ``screencast, record screen as video. Screen capture for full page,,annotate, blur sensitive info, and share with one-click uploads" & Browsing History          \\
Hover Zoom                                & hoverzoom                  & 894,888    & ``enlarge thumbnails on mouse over. Works on many sites (Facebook, Twitter, Flickr, Reddit, Amazon, Tumblr, etc)"                   & Browsing History          \\
SpeakIt                                & skechboy.com                  & 545,715    & ``Select text you want to read and listen to it. SpeakIt converts text into speech so you no longer need to read."                   & Browsing History          \\
MozBar                                    & Moz                     & 375,529    & ``the all-in-one SEO toolbar for research on the go."                                                                               & Browsing History          \\ 
User-Agent Switcher                       & useragentswitcher          & 299,771    & ``...adds a button to switch between user-agents..."                                                                                & Browsing History          \\
Facebook Video Downloader                   & freevideodownloader & 272,275    & ``download any video from Facebook"                                                                                                 & Browsing History \\
Ratings Preview for YouTube               & ratingspreview             & 116,745    & ``show the likes and dislikes bar over every video thumbnail in YouTube"                                                            & Browsing History          \\
SafeBrowse                                & safebrowse.co              & 115,265    & ``navigate without waiting in Adfly, Linkbucks and other similar sites. No more waiting, no more annoying advertising"              & Browsing History          \\
SuperBlock Adblocker                      & Superblock                 & 13,887    & Ad blocker                                                                                                                          & Browsing History          \\
Power Zoom                                & powerzoom                  & --    & ``to view larger images on any website automatically, all over the Web"                                                             & Browsing History          \\
\bottomrule
\end{tabular}
\caption{A sample of newly found spying extensions.}
\label{tab:falsepos}
\end{table*}

\end{document}